\documentclass[a4paper,draft]{amsart}
\usepackage[a4paper,margin=25mm]{geometry}
\usepackage{amsmath,amssymb,url,longtable}

\parskip\smallskipamount

\let\set\mathbb
\def\<#1>{\langle#1\rangle}

\def\OO{\mathrm{O}}

\begin{document}

 \allowdisplaybreaks

 \author[Manuel Kauers and Jakob Moosbauer]{Manuel Kauers\,$^\ast$ and Jakob Moosbauer\,$^\dagger$}
 \address{Manuel Kauers, Institute for Algebra, J. Kepler University Linz, Austria}
 \email{manuel.kauers@jku.at}
 \address{Jakob Moosbauer, Institute for Algebra, J. Kepler University Linz, Austria}
 \email{jakob.moosbauer@jku.at}
 \thanks{$^\ast$ Supported by the Austrian FWF grant P31571-N32.\\
         $^\dagger$ Supported by the Land Oberösterreich through the Lit-AI Lab.}

 \title{The FBHHRBNRSSSHK-Algorithm for multiplication in $\set Z_2^{5\times5}$ is still not the end of the story}

 \begin{abstract}
   In response to a recent \emph{Nature} article which announced an algorithm for multiplying $5~\times~5$-matrices
   over $\set Z_2$ with only 96 multiplications, two fewer than the previous record,
   we present an algorithm that does the job with only 95 multiplications.
 \end{abstract}

 \maketitle

 \section{Introduction}

 Ever since Strassen~\cite{strassen1969gaussian} discovered that $2\times 2$-matrices can be multiplied with only 7~multiplications
 in the coefficient domain, there is a mystery around the complexity of matrix multiplication.
 For asymptotically large~$n$, the best we know at the moment is a multiplication algorithm that requires $\OO(n^{2.3728596})$
 operations~\cite{alman20}, slightly improving upon the previous record~$\OO(n^{2.3728639})$~\cite{LeGall:2014:PTF:2608628.2608664}.
 For $n=3$, it is known that 23~multiplications suffice in a non-commutative setting~\cite{laderman1976noncommutative}.
 For $n=4$, we can solve the problem with 49~multiplications by applying Strassen's algorithm recursively.
 In a recent article that received considerable media attention, Fawzi et al.~\cite{fawzi22} used a machine learning
 approach to find a multiplication scheme with 47 multiplications, applicable to coefficient domains of
 characteristic~$2$. Under the same restriction on the coefficient domain, they also improved the best known
 bound for $n=5$ from 98~\cite{sedoglavic21} to~96.
 See \cite{fawzi22,fastmm} for current records for other formats and further references on the matter.

 In this short note, we present another non-equivalent solution for $4\times4$-matrices requiring 47 multiplications,
 as well as a first solution for $5\times5$-matrices requiring 95 multiplications.
 This solution was obtained from the scheme of Fawzi et al. by applying a sequence of transformations
 leading to a scheme from which one multiplication could be eliminated.
 Our new scheme for $4\times4$-matrices was obtained by the same technique, taking the standard
 multiplication algorithm (with 64 multiplications) as starting point.
 We will explain our search technique in further detail in a forthcoming paper~\cite{moosbauer22}.
 
 \section{Multiplication of $4\times4$ Matrices}

 Let
 \[
 A=
:=AB
 \]
 be matrices with coefficients in a ring $R$ of characteristic~2.

 We can then compute the entries of $C$ from the entries of $A$ and $B$ by first setting
 \[
   m_k = \biggl(\sum_{i=1}^4\sum_{j=1}^4\alpha^{(k)}_{i,j}a_{i,j}\biggr)\biggl(\sum_{i=1}^4\sum_{j=1}^4\beta^{(k)}_{i,j}b_{i,j}\biggr)
 \]
 for $k=1,\dots,47$ and then
 \[
   c_{i,j} = \sum_{k=1}^{47}\gamma^{(k)}_{i,j}m_k
 \]
 for $i,j=1,\dots,4$, where the $\alpha^{(k)}_{i,j},\beta^{(k)}_{i,j},\gamma^{(k)}_{i,j}$ are as follows:

 \begin{longtable}{c|c|c|c}
   $k$ & $((\alpha^{(k)}_{i,j}))_{i,j=1}^4$ & $((\beta^{(k)}_{i,j}))_{i,j=1}^4$ & $((\gamma^{(k)}_{i,j}))_{i,j=1}^4$ \\\hline
1&$\begin{pmatrix} 0 & 0 & 1 & 0 \\ 1 & 1 & 0 & 0 \\ 0 & 1 & 0 & 0 \\ 1 & 1 & 1 & 0 \end{pmatrix}$&$\begin{pmatrix} 0 & 1 & 0 & 1 \\ 0 & 1 & 0 & 1 \\ 0 & 0 & 0 & 0 \\ 0 & 1 & 0 & 1 \end{pmatrix}$&$\begin{pmatrix} 0 & 0 & 0 & 0 \\ 0 & 1 & 0 & 0 \\ 0 & 1 & 0 & 0 \\ 0 & 0 & 0 & 0 \end{pmatrix}$\\\hline
2&$\begin{pmatrix} 1 & 0 & 1 & 0 \\ 1 & 1 & 0 & 0 \\ 1 & 1 & 0 & 0 \\ 1 & 1 & 1 & 0 \end{pmatrix}$&$\begin{pmatrix} 0 & 1 & 1 & 0 \\ 0 & 0 & 0 & 0 \\ 0 & 1 & 0 & 1 \\ 0 & 1 & 1 & 0 \end{pmatrix}$&$\begin{pmatrix} 0 & 1 & 0 & 0 \\ 0 & 1 & 0 & 0 \\ 0 & 0 & 0 & 0 \\ 0 & 1 & 0 & 0 \end{pmatrix}$\\\hline
3&$\begin{pmatrix} 0 & 1 & 1 & 0 \\ 0 & 0 & 0 & 0 \\ 0 & 0 & 0 & 0 \\ 1 & 0 & 0 & 1 \end{pmatrix}$&$\begin{pmatrix} 0 & 0 & 1 & 1 \\ 0 & 0 & 0 & 0 \\ 0 & 0 & 0 & 0 \\ 0 & 1 & 0 & 0 \end{pmatrix}$&$\begin{pmatrix} 0 & 1 & 1 & 0 \\ 0 & 1 & 1 & 0 \\ 0 & 0 & 0 & 1 \\ 0 & 1 & 1 & 1 \end{pmatrix}$\\\hline
4&$\begin{pmatrix} 1 & 1 & 0 & 0 \\ 1 & 1 & 0 & 0 \\ 0 & 0 & 0 & 0 \\ 0 & 0 & 0 & 0 \end{pmatrix}$&$\begin{pmatrix} 0 & 0 & 0 & 1 \\ 1 & 0 & 1 & 1 \\ 0 & 0 & 0 & 0 \\ 1 & 1 & 1 & 1 \end{pmatrix}$&$\begin{pmatrix} 0 & 1 & 1 & 0 \\ 0 & 1 & 0 & 0 \\ 0 & 0 & 0 & 1 \\ 0 & 1 & 0 & 1 \end{pmatrix}$\\\hline
5&$\begin{pmatrix} 0 & 1 & 1 & 1 \\ 0 & 0 & 0 & 0 \\ 0 & 0 & 0 & 0 \\ 0 & 0 & 0 & 0 \end{pmatrix}$&$\begin{pmatrix} 1 & 0 & 0 & 0 \\ 0 & 0 & 0 & 0 \\ 1 & 0 & 1 & 1 \\ 0 & 1 & 1 & 0 \end{pmatrix}$&$\begin{pmatrix} 1 & 0 & 0 & 0 \\ 1 & 0 & 0 & 0 \\ 0 & 0 & 0 & 0 \\ 1 & 0 & 0 & 0 \end{pmatrix}$\\\hline
6&$\begin{pmatrix} 0 & 0 & 0 & 0 \\ 0 & 1 & 0 & 0 \\ 0 & 1 & 0 & 0 \\ 0 & 1 & 0 & 0 \end{pmatrix}$&$\begin{pmatrix} 1 & 1 & 1 & 1 \\ 1 & 1 & 1 & 1 \\ 0 & 1 & 0 & 0 \\ 1 & 1 & 1 & 1 \end{pmatrix}$&$\begin{pmatrix} 0 & 0 & 0 & 0 \\ 0 & 1 & 0 & 0 \\ 0 & 1 & 1 & 1 \\ 1 & 1 & 0 & 1 \end{pmatrix}$\\
\end{longtable}

 An electronic version of this multiplication scheme is available at
 \url{http://www.algebra.uni-linz.ac.at/people/mkauers/matrix-mult/s47.exp}
 
 \section{Multiplication of $5\times5$ Matrices} 
 
 Let
 \[
 A=%
:=AB
 \]
 be matrices with coefficients in a ring $R$ of characteristic~2.

 We can then compute the entries of $C$ from the entries of $A$ and $B$ by first setting
 \[
   m_k = \biggl(\sum_{i=1}^5\sum_{j=1}^5\alpha^{(k)}_{i,j}a_{i,j}\biggr)\biggl(\sum_{i=1}^5\sum_{j=1}^5\beta^{(k)}_{i,j}b_{i,j}\biggr)
 \]
 for $k=1,\dots,95$ and then
 \[
   c_{i,j} = \sum_{k=1}^{95}\gamma^{(k)}_{i,j}m_k
 \]
 for $i,j=1,\dots,5$, where the $\alpha^{(k)}_{i,j},\beta^{(k)}_{i,j},\gamma^{(k)}_{i,j}$ are as follows:

 \begin{longtable}{c|c|c|c}
   $k$ & $((\alpha^{(k)}_{i,j}))_{i,j=1}^5$ & $((\beta^{(k)}_{i,j}))_{i,j=1}^5$ & $((\gamma^{(k)}_{i,j}))_{i,j=1}^5$ \\\hline
1&$\begin{pmatrix} 0 & 0 & 0 & 0 & 0 \\ 0 & 1 & 0 & 0 & 0 \\ 0 & 0 & 0 & 0 & 0 \\ 0 & 0 & 0 & 0 & 0 \\ 0 & 0 & 0 & 0 & 0\end{pmatrix}$&$\begin{pmatrix} 0 & 0 & 0 & 0 & 0 \\ 0 & 0 & 0 & 1 & 0 \\ 0 & 0 & 0 & 0 & 0 \\ 0 & 0 & 0 & 0 & 0 \\ 0 & 0 & 0 & 0 & 0 \end{pmatrix}$&$\begin{pmatrix} 0 & 0 & 0 & 0 & 0 \\ 1 & 0 & 0 & 1 & 0 \\ 0 & 0 & 0 & 0 & 0 \\ 0 & 0 & 0 & 1 & 0 \\ 0 & 0 & 0 & 0 & 0 \end{pmatrix}$\\\hline
2&$\begin{pmatrix} 0 & 0 & 0 & 0 & 0 \\ 0 & 1 & 0 & 0 & 0 \\ 0 & 0 & 0 & 0 & 0 \\ 0 & 1 & 1 & 0 & 0 \\ 0 & 0 & 0 & 0 & 0\end{pmatrix}$&$\begin{pmatrix} 0 & 0 & 0 & 0 & 0 \\ 0 & 0 & 1 & 0 & 0 \\ 0 & 0 & 1 & 1 & 0 \\ 0 & 0 & 0 & 0 & 0 \\ 0 & 0& 0 & 0 & 0 \end{pmatrix}$&$\begin{pmatrix} 0 & 0 & 0 & 0 & 0 \\ 0 & 0 & 1 & 0 & 0 \\ 0 & 0 & 0 & 0 & 0 \\ 0 & 0 & 0 & 1 &0 \\ 0 & 0 & 0 & 0 & 0 \end{pmatrix}$\\\hline
3&$\begin{pmatrix} 0 & 0 & 0 & 0 & 0 \\ 0 & 1 & 0 & 0 & 0 \\ 0 & 0 & 0 & 0 & 0 \\ 0 & 1 & 0 & 0 & 1 \\ 0 & 0 & 0 & 0 & 0\end{pmatrix}$&$\begin{pmatrix} 0 & 0 & 0 & 0 & 0 \\ 0 & 1 & 0 & 0 & 0 \\ 0 & 0 & 0 & 0 & 0 \\ 0 & 0 & 0 & 0 & 0 \\ 0 & 1& 0 & 1 & 0 \end{pmatrix}$&$\begin{pmatrix} 0 & 0 & 0 & 0 & 0 \\ 0 & 1 & 0 & 0 & 0 \\ 0 & 0 & 0 & 0 & 0 \\ 0 & 0 & 0 & 1 &0 \\ 0 & 0 & 0 & 0 & 0 \end{pmatrix}$\\\hline
4&$\begin{pmatrix} 0 & 0 & 0 & 0 & 0 \\ 0 & 1 & 0 & 0 & 0 \\ 0 & 0 & 0 & 0 & 0 \\ 1 & 1 & 0 & 0 & 0 \\ 0 & 0 & 0 & 0 & 0\end{pmatrix}$&$\begin{pmatrix} 0 & 0 & 0 & 1 & 1 \\ 0 & 0 & 0 & 0 & 1 \\ 0 & 0 & 0 & 0 & 0 \\ 0 & 0 & 0 & 0 & 0 \\ 0 & 0& 0 & 0 & 0 \end{pmatrix}$&$\begin{pmatrix} 0 & 0 & 0 & 0 & 0 \\ 0 & 0 & 0 & 0 & 1 \\ 0 & 0 & 0 & 0 & 0 \\ 0 & 0 & 0 & 1 &0 \\ 0 & 0 & 0 & 0 & 0 \end{pmatrix}$\\
\hline
 \end{longtable}

 An electronic version of this multiplication scheme is available at
 \url{http://www.algebra.uni-linz.ac.at/people/mkauers/matrix-mult/s95.exp}
 
 \bibliographystyle{plain}
 \bibliography{main}
 
\end{document}